\documentstyle[11pt,epsfig]{article}

\fussy
\def\be{\begin{equation}}
\def\ee{\end{equation}}
\def\bea{\begin{eqnarray}}
\def\eea{\end{eqnarray}}

\def\la{\mathrel{\mathpalette\fun <}}
\def\ga{\mathrel{\mathpalette\fun >}}
\def\fun#1#2{\lower3.6pt\vbox{\baselineskip0pt\lineskip.9pt
\ialign{$\mathsurround=0pt#1\hfil##\hfil$\crcr#2\crcr\sim\crcr}}}  

\begin{document}

\begin{titlepage} 

\date{ } 

\vskip 4 cm 

\title{\bf IN-MEDIUM PARTON ENERGY LOSSES AND CHARACTERISTICS OF HADRONIC JETS IN
ULTRA-RELATIVISTIC NUCLEAR COLLISIONS} 

\vskip 1cm 

\author{\Large I.P.Lokhtin \\ ~ \\ 
\it 119899, Moscow State University, Nuclear Physics Institute, Moscow, Russia}

\maketitle 

\vskip 2cm   

\begin{abstract} 

The angular structure of collisional and radiative energy losses of a hard parton jet 
propagating through the quark-gluon plasma is analyzed. The possibility to observe the 
energy losses of quark- and gluon-initiated jets in dense QCD-matter (jet quenching) 
measuring the characteristics of real hadronic jets in ultra-relativistic collisions of 
nuclei is investigated. In particular, using calorimetric studies of $jet + jet$, 
$\gamma + jet$ and $Z + jet$ channels is discussed.  

\vskip 8cm  

\noindent 
\underline{\hspace{8cm}} \\ 
Talk given at XXXIVth Rencontres de Moriond \\  
"QCD and High Energy Hadronic Interactions", \\ 
Les Arcs, France, March 20-27, 1999

\end{abstract} 
\end{titlepage}
\section{Angular structure of energy losses of hard jet in dense QCD-matter}

Hard jet production is considered to be an efficient probe for formation of 
quark-gluon plasma (QGP)~\cite{qm97} in future experiments on 
heavy ion collisions at LHC~\cite{cms94,alice}. High $p_T$ parton pair (dijet) from 
a single hard scattering is produced at the initial stage of the collision process 
(typically, at $\la 0.01$ fm/c). It then propagates through the QGP formed due to 
mini-jet production at larger time scales ($\sim 0.1$ fm/c), and interacts strongly 
with the comoving constituents in the medium. 

We know two possible mechanisms of energy losses of a hard partonic jet evolving 
through the dense matter : $(1)$ radiative losses due to gluon "bremsstrahlung" induced 
by multiple scattering~\cite{gyul94,baier,zakharov} and $(2)$ collisional losses 
due to the elastic rescatterings of high $p_T$ partons off the medium 
constituents~\cite{mrow91,lokhtin1}. Although the radiative energy losses of a 
high energy parton can dominate over the collisional losses by up to an order of 
magnitude~\cite{gyul94}, the angular distribution of the losses is essentially 
different for two mechanisms. Indeed, the coherent Landau-Pomeranchuk-Migdal radiation 
induces a strong dependence of the jet energy on the jet cone size 
$\theta_0$~\cite{baier,lokhtin2}. With increasing of hard parton energy the maximum of 
the angular distribution of bremsstrahlung gluons shifts towards the parent parton 
direction. This means that measuring the jet energy as a sum of the energies of final 
hadrons moving inside an angular cone with a given finite size $\theta_0$ will allow 
the bulk of the gluon radiation to belong to the jet. Therefore, the medium-induced 
radiation will, in the first place, soften particle energy distributions inside the 
jet, increase the multiplicity of secondary particles, but will not affect the total 
jet energy. On the other hand, the collisional energy losses turns out to be 
practically independent on $\theta_0$ and emerges outside the narrow jet cone: the bulk 
of "thermal" particles knocked out of the dense matter by elastic scatterings fly away 
in almost transverse direction relative to the jet axis. 

The total energy loss experienced by a hard parton due to multiple 
scattering in matter is the result of averaging over the dijet production vertex 
($R$, $\varphi$), the momentum transfer $t$ in a single rescattering  
and space-time evolution of the medium:  
\begin{equation}
\Delta E_{tot} = 
\int\limits_0^{2\pi}\frac{d\varphi}{2\pi}\int\limits_0^{R_A}dR\cdot P_A(R)
\int\limits_{\displaystyle\tau_0}^{\displaystyle 
\tau_L}d\tau \left( \frac{dE}{dx}^{rad}(\tau) + \sum_{b}\sigma_{ab}(\tau)\cdot
\rho_b(\tau)\cdot \nu(\tau) \right) .    
\end{equation} 
Here $\tau_0$ and $\tau_L = \sqrt{R_A^2-R^2\sin^2{\varphi}} - R\cos{\varphi}$ are the 
proper time of the QGP formation and the time of jet escaping from the plasma;    
$P_A(R)$ is the distribution of the distance $R$ from the axis of nuclei collision $z$ 
to the dijet production vertex; 
$n_b \propto T^3$ is the density of plasma constituents of type $b$ at temperature $T$; 
$\sigma_{ab}$ is the integral cross section of scattering of a jet parton $a$ off the 
comoving constituent $b$.   
$\frac{dE}{dx}^{rad}(\tau)$ is the 
radiative energy losses per unit length; $\nu(\tau) = <Q^2 / 2m_0>$ is the thermal 
average collisional energy loss of the jet parton with energy $E$ due to elastic 
single scattering off a constituent of the medium with energy $m_0 \sim 3T$. The 
scatterings can be treated as independent, if the mean free path of hard parton is 
larger than the Debye screening radius, $\lambda \equiv \sum_{b} \sigma_{ab} n_b \gg 
\mu_D^{-1}$. 

We have suggested a simple generalization of BDPMS result~\cite{baier} for 
$dE / dx^{rad}$ to calculate
the gluon energy deposited outside a given cone $\theta_0$, which  
is based on the relation between the gluon radiation angle $\theta$ and energy 
$\omega$ which holds only in {\it average}~\cite{lokhtin2}:  
\begin{equation} \label{angen}
  \bar{\theta}=\theta(\omega) 
  \simeq \theta_M\cdot \left(\frac{E_{LPM}}{\omega}\right)^\frac34, 
\end{equation}
where $E_{LPM}=\mu_D^2\lambda_g$ is the minimal radiated gluon energy in the 
coherent LPM regime and $\theta_M \>=\> (\mu_D\lambda_g)^{-1}\>$ is the characteristic 
angle depending on the local properties of the medium. 
Note that the problem of a rigorous description of the differential angular 
(transverse momentum) distribution of induced radiation is complicated 
by intrinsically quantum-mechanical nature of the phenomenon: large
formation times of the radiation does not allow the direction of the
emitter to be precisely defined~\cite{baier}. 

\parindent 0mm
\vspace{6mm}
%\begin{minipage}[1]{80mm}
%\begin{center} 
\epsfig{file=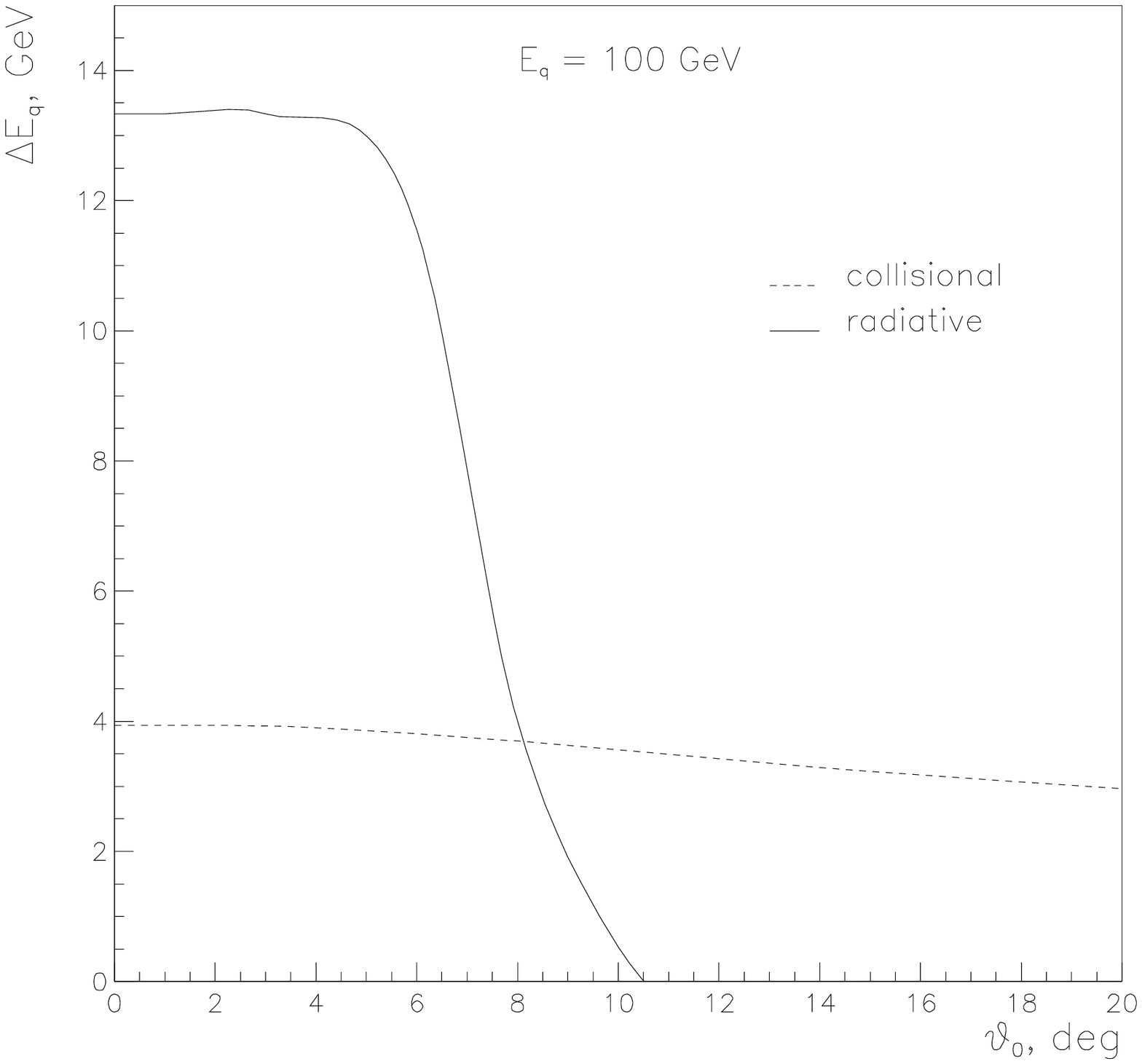,width=70 mm}
%\end{center}
\begin{figure}[htbp]
\end{figure}
%\end{minipage}

\vspace{-15mm}
\begin{small}
Figure 1: ~The average radiative ~(solid curve) 

and collisional (dashed curve) energy losses of 

quark-initiated ~jet ~as ~a ~function ~of ~the jet 

cone size $\theta_0$

\end{small}

\vspace{-83mm}
\hspace{78mm}
\begin{minipage}{81mm}

Figure 1 represents the average radiative (coherent medium-dependent part) and 
collisional energy losses of a quark-initiated jet with initial energy $E = 100$ GeV  
as a function of the jet cone size $\theta_{0}$. We have used scaling Bjorken's 
solution~\cite{bjorken} for temperature and density of gluon-dominated plasma  
$T(\tau) \tau^{1/3} = T_0 \tau_0^{1/3},~~ \rho(\tau) \tau = \rho_0 \tau_0$ and initial
conditions, predicted to be achieved in central $Pb-Pb$ collisions at LHC 
energies~\cite{eskola94}: $\tau_0 \simeq 0.1$ fm/c, $T_0 \simeq 1$ GeV.  

We can see the weak $\theta_0$-dependence of collisional losses, 
at least $90 \%$ of scattered ``thermal'' particles 
flow outside a rather wide cone $\theta_0 \sim 10^0 - 20^0$. 
The radiative losses are almost independent of the initial jet energy and 
decrease rapidly with increasing the angular size of the jet at $\theta_0 \ga 5^0$. 

\end{minipage}

\bigskip
 
\section{Jet quenching: $jet + jet$, $\gamma + jet$ and $Z + jet$ production}

In a search for experimental evidences in favour of the medium-induced 
energy losses a significant dijet quenching (a suppression of high $p_T$ jet
pairs)~\cite{gyul90} was proposed as a possible signal of dense matter formation in 
ultra-relativistic nuclear collisions. Note that the dijet rate in $AA$ relative to $pp$ 
collisions can be studied 
by introducing a reference process, unaffected by energy losses and with a rate 
proportional to the number of nucleon-nucleon collisions, such as Drell-Yan dimuons or 
Z$(\to \mu^+\mu^-)$ production:  
$ 
R^{dijet}_{AA} / R^{dijet}_{pp} = \left( \sigma_{AA}^{dijet} / 
\sigma_{pp}^{dijet} \right) /  \left(\sigma_{AA}^{DY~(Z)} / \sigma_{pp}^{DY~(Z)} 
\right).   
$ 

We have studied the capability of the CMS detector~\cite{cms94} at future LHC collider 
to observe the medium-induced energy losses of quarks and gluon detecting   
hadronic jets in heavy ion collisions~\cite{note99_016}. The Compact Muon Solenoid 
(CMS) is the general purpose detector designed to run at the LHC and optimized mainly 
for the search of the Higgs boson in $pp$ collisions. However, a good muon system and 
electromagnetic and hadron calorimeters with fine granularity gives the possibility to 
cover important "hard probes" aspects of the heavy ion physics. 
At LHC ions will be accelerated at $\sqrt{s} = 7 \times (2Z/A)$ TeV per 
nucleon pair. In the case of $Pb$ nuclei $\sqrt{s} = 5.5 A$
TeV and the expected average luminosity $L \approx 1.0 \times 
10^{27}$~cm$^{-2}$s$^{-1}$. The inelastic interaction cross-section for $Pb-Pb$ 
collisions is about 8~b, which leads to an event rate of 8~kHz. 

The jet recognition efficiency and expected production rates was studied assessing the 
CMS calorimeters response for the barrel part of CMS calorimeters, which covers the 
pseudorapidity region of $\mid \eta \mid < 1.5$. Using the selection criterion on jet 
shape allows getting the maximum efficiency of "true" hard jets 
recognition as well as the maximum suppression of "false" jets background at jet
energy $E_T \sim 50-100$ GeV~\cite{note99_016}. In order to test the sensitivity of 
the final hadronic jets to the energy losses, the three different 
scenarios for jet quenching due to collisional energy losses of a jet  
partons were studied~\cite{lokhtin1}: $(i)$ no jet quenching,
$(ii)$ jet quenching in a perfect QGP (the average collisional losses 
of a hard gluon $<\Delta E_{g}> \simeq 9$ GeV, $<\Delta E_{q}> = 4/9 
\cdot <\Delta E_{g}>$), $(iii)$ jet quenching in a maximally viscous QGP,
resulting in  $<\Delta E_{g}> \simeq 18$ GeV. 

\vspace{6mm}
\hspace{3mm}
\epsfig{file=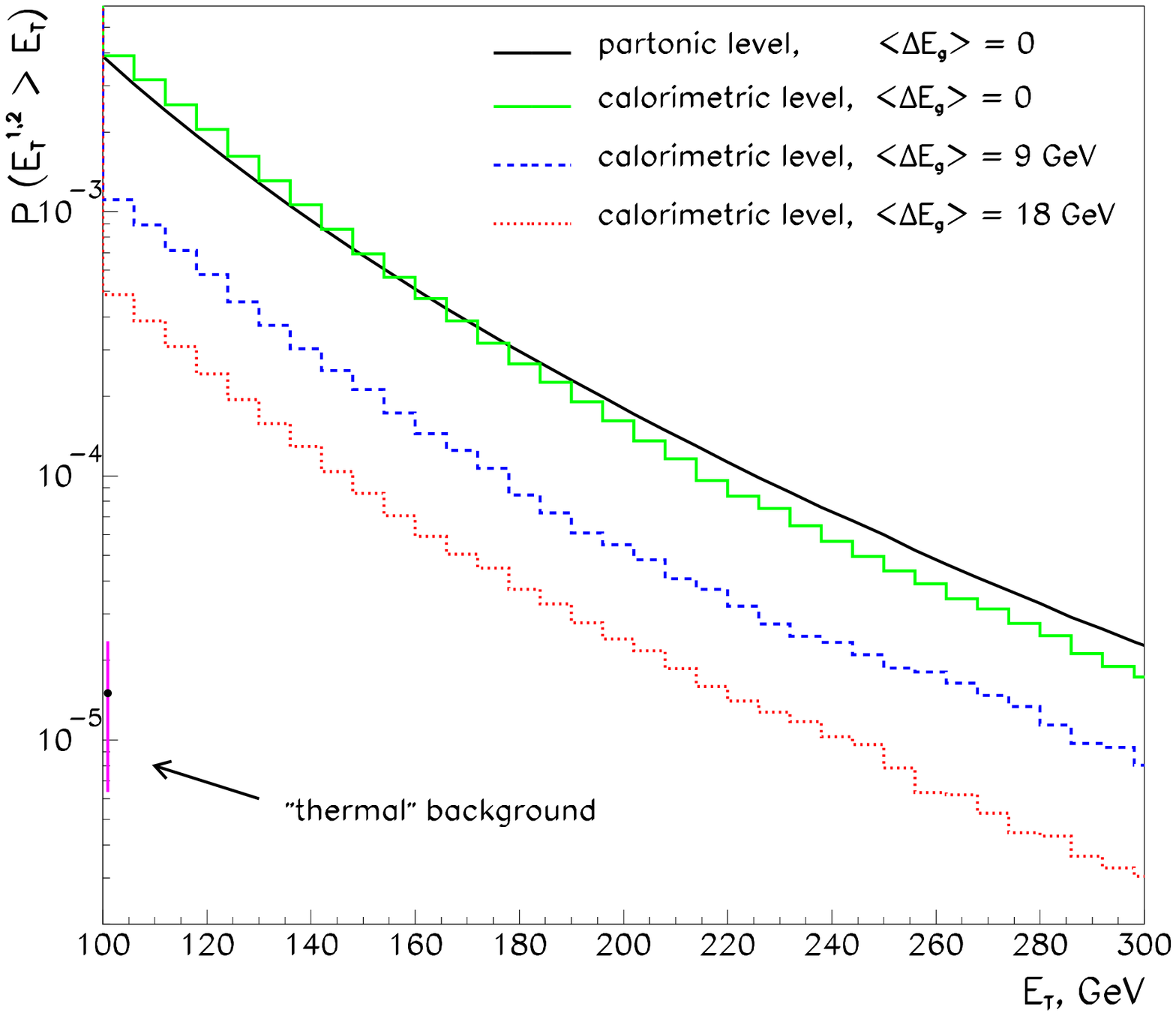,height=70mm}
\begin{figure}[hbtp]
\end{figure}

\vspace{-4mm}
\begin{small}
Figure 2: ~The probability ~$P_{dijet}$ ~of dijet production ~with 

transverse energy $E_T^{jet1,2} > E_T$ in central $Pb-Pb$ collisions 

for different quenching scenarios: ~"true" hard ~(histograms) 

and ~"false" (point) ~dijets, ~$dN^{\pm}/dy (y=0) = 8000$. ~Solid 

curve shows ~the scaled result ~for dijet spectrum ~at parton 

level as calculated with PYTHIA.
\end{small}

\vspace{-106mm}
\hspace{98mm}
\begin{minipage}{61mm}

Figure 2 represents the probability of dijet yield as a function of  
$E_T$ in simulated central $Pb-Pb$ collisions for CMS. 
The significant suppression of hard dijet yield due to energy losses (up to factor 
$\sim 7$, or somewhat higher if the radiative energy losses mechanism is included) can 
be expected. The quenching factor is almost independent of the jet energy if the 
losses do not depend (or depend weakly) on the energy of an initial hard parton. 
The expected statistics for dijet production will be large 
enough to make the study the dijet rates as a function of impact parameter of the 
collision and the transverse energy of jets. The suppression of dijet rates (jet 
quenching) due to energy losses of hard partons expected to be much more stronger at 
very central collisions in comparison with the peripheral one's. 

\end{minipage}

\vspace{10mm}

Other possible signatures that could directly observe the energy losses involve
tagging the hard jet opposite a particle that does not interact strongly
($q + g \rightarrow q + \gamma$~\cite{wang96} and $q + g \rightarrow q + 
Z(\to \mu^+\mu^-)$~\cite{kvat95} channels). The jet energy losses
should result in the non-symmetric shape of the distribution of differences in 
transverse momentum between the Z-boson ($\gamma$) and jet. The estimated statistics is 
rather low for Z$(\to \mu^+\mu^-)$ + jet channel. On the other hand, using 
$\gamma + jet$ production is complicated due to large background from $jet+jet$ 
production when one of the jet in an event is misidentified as a photon (the leading 
$\pi ^{0}$). However the shape of the distribution of differences in transverse energy 
between the $\gamma$ and jet is well sensitive to the jet quenching effect. 
It seems possible to extract the background $\gamma + jet$ events from the experimental 
spectra using the background shape from Monte-Carlo simulation and (or) from $pp$ data. 

\section{Conclusions} 
For small angular jet cone sizes, $\theta_0\la 5^0$,  
the radiative energy loss is shown to dominate over the collisional
energy loss due to final state elastic rescattering of the hard
projectile on thermal particles in the medium. Due to coherent 
effects, the radiative energy loss decreases with increasing the
angular size the jet. It becomes comparable with the collisional energy
loss for $\theta_0 \ga 5^0-10^0$. Relative 
contribution of collisional losses would likely become significant for jets with finite 
cone size propagating through the hot plasma under LHC conditions

Monte-Carlo study shows that CMS detector at future LHC collider is well suited for 
the investigation of high transverse energy jets. Dijet production, 
$Z(\to \mu^+\mu^-) + jet$ and $\gamma + jet$ channels are important for 
extracting information about the properties of super-dense matter to be created in 
heavy ion collisions at LHC.

\section*{Acknowledgments}
I would like to thank A.Nikitenko, L.I.Sarycheva, A.M.Snigirev and 
I.N.Vardanian for collaboration on parts of this work. Discussions with M.Bedjidian,
D.Denegri, Yu.L.Dokshitzer  O.Drapier, O.L.Kodolova, R.Kvatadze and D.Schiff are 
gratefully acknowledged.

\end{document}